\documentclass{article}
\begin{document}
\begin{titlepage}
\title{The Hamiltonian of Asymptotically
Friedmann-Lema\^{\i}tre-Robertson-Walker Spacetimes}
\author{N. Pinto-Neto \footnote{Laborat\'orio de Cosmologia e
F\'{\i}sica Experimental de Altas
Energias, Centro Brasileiro de Pesquisas Fisicas,
Rua Dr. Xavier
Sigaud 150, Urca, Rio de Janeiro CEP 22290-180-RJ, Brazil.}
\footnote{e-mail address: nelsonpn@cbpf.br}
and P. I. Trajtenberg${}^{\ast}$}
\maketitle
\begin{abstract}
We obtain the correct hamiltonian which describes the dynamics
of classes of asymptotic open Friedmann-Lema\^{\i}tre-Robertson-Walker (FLRW)
spacetimes, which includes Tolman geometries. We calculate the
surface term that has to be added to the usual hamiltonian of General Relativity
in order to obtain an improved hamiltonian with well defined
functional derivatives. For asymptotic flat FLRW spaces, this surface term
is zero, but for asymptotic negative curvature FLRW spaces it is not
null in general. In the particular case of the Tolman geometries, they vanish.
The surface term evaluated on a particular solution
of Einstein's equations may be viewed as the ``energy'' of this solution
with respect to the FLRW spacetime they approach
asymptotically. Our results are obtained for a matter content
described by a dust fluid, but they are valid for any perfect
fluid, including the cosmological constant.

\vspace{0.5cm}
PACS number(s): 04.20.Cv., 04.20.Me

\vspace{0.5cm}
\end{abstract}
\end{titlepage}

\section{Introduction}

The hamiltonian formulation of General Relativity (GR) is known
to have very different properties depending whether it is realized in
closed spaces or in open spaces. For closed spaces, the hamiltonian
of GR can be written in terms of a single general expression consisting
of a volume integral in a spacelike hypersurface whose integrand is a
hamiltonian density which is a combination of constraints \cite{dir}.
For open spaces, however, it was long ago recognized that the hamiltonian
of GR must be suplemented by a surface term \cite{dir}. In Ref.\cite{tei1},
 it was shown that such surface terms are in fact necessary if
one wants to construct a consistent hamiltonian of GR in open spaces,
with well defined functional derivatives. The specific form of
the surface term depends on the asymptotic structure of such spaces.
In this way, the surface term of Ref.\cite{dir} was obtained in Ref.\cite{tei1}
from a consistency criterium applied to asymptotically flat spacetimes.
This procedure was then used in asymptotically anti-de Sitter (ADS)
spacetimes \cite{tei2} in order to obtain the correct hamiltonian of
such spaces.

The evaluation of the total hamiltonian of GR on shell can be viewed
as the total ``energy'' of specific gravitational fiels solving the Einstein's
equations. For closed spaces this value is zero (any solution of Einstein's
equations must, of course, satisfy the constraints). For open spaces
with asymptotically flat and ADS structures, this value is given by the
surface term evaluated on shell, which is not zero. The results are in
agreement with other prescriptions to evaluate the total energy of these
gravitational fields \cite{pseudoflat,pseudoADS}
(see however \cite{iva}).

The aim of this paper is to apply the procedure described in
Refs.\cite{tei1,tei2}
to more involved asymptotic structures. As we live in a Universe very well
described by Friedmann-Lema\^{\i}tre-Robertson-Walker (FLRW) spacetimes,
we will study the hamiltonian formulation of asymptotically FLRW spacetimes.
The matter content will be described by a dust fluid, but, as we will see,
the results we obtain are valid for any perfect fluid, including the cosmological
constant.
The asymptotic structure of such spaces was obtained as follows: we
started with Tolman geometries which are asymptotically FLRW spaces, and we
act on them with the group of isometries of FLRW spaces in all possibles ways,
obtaining a class of asymptotic FLRW geometries which contains Tolman
geometries and which are invariant under the action of the Killing vectors
of FLRW spaces at infinity. This procedure yields the asymptotic structure of
the 3-geometries
and their momenta. The asymptotic behaviors of the lapse and shift functions are
obtained through the requirements that the hamiltonian dynamics preserves
the asymptotic structure of the spatial geometries, and that the fluid
4-velocities
be normal to the spacelike hypersurfaces which folliate spacetime.
Of course, this is not the most general class of asymptotic FLRW spaces, but
it contains a wide variety of physically interesting solutions such as Tolman
geometries.

This paper is organized as follows: in the next section we review the
main general aspects of the hamiltonian formulation of GR for open spaces.
In section III we describe in details the construction of the class
of asymptotic FLRW spaces we will work with. In section IV we arrive at
the correct hamiltonian which describe the dynamics of such solutions.
We end up with the conclusions.

\section{The Hamiltonian of General Relativity For Open Spaces}

The hamiltonian of General Relativity for closed geometries and a dust
field is given by
\begin{equation}
\label{ham}
H = \int{N_{\mu}{\cal{H}}^{\mu}{\rm d}^{3}x}
\end{equation}
where $N_{\mu} = (N , N_{i})$, $N$ and $N^i$ are the lapse and shift functions,
respectively, and $\cal{H}^{\mu}$ is given by
\begin{equation}
\begin{array}{l}
{\cal{H}}^{0} = 16\pi G_{abcd}\pi^{ab}\pi^{cd} - \frac{\sqrt{g}R}{16\pi} +
(m^{2} + \chi_{, i}\chi^{, i})^{1/2}\pi_{\chi}, \\
{\cal{H}}^{i} = - 2\pi^{ik}_{\| k} + \pi_{\chi}\chi^{, i},
\end{array}
\end{equation}
with $G_{abcd} = \frac{1}{2}g^{-1/2}[g_{ac}g_{bd} + g_{ad}g_{bc} - g_{ab}g_{cd}]$.
In the above expressions $g_{ik}$ is the metric, $g$ its determinant,
and $R$ the scalar curvature of spacelike hypersurfaces. The
canonical gravitational momentum is given by $\pi^{ik}$, the double bar
represents the covariant
derivative with respect to $g_{ik}$, and $\chi$ is the dust field defining
the velocity field of the dust particles of mass $m$,
\begin{equation}
\label{velocity}
u^{\mu} = - \frac{{}^{4}g^{\mu\nu}\chi_{, \nu}}{m}.
\end{equation}
where ${}^{4}g^{\mu\nu}$ is the inverse of the 4 - dimensional metric
${}^{4}g_{\mu\nu}$
The quantity $\pi_{\chi}$ is the canonical momentum of $\chi$,
which is connected to the number density of dust paticles $n$ through
\begin{equation}
\label{momd}
\pi_{\chi} = ng^{1/2}\biggl(1 + \frac{\chi_{, i}\chi^{, i}}{m^{2}}\biggr).
\end{equation}
We are using geometrical units where $G=c=1$.

When one varies H with respect to $g_{ik}$ and $\pi^{ik}$, $\chi$ and
$\pi_{\chi}$ one obtains:
\begin{equation}
\label{deltaH}
\delta H = \int{d^{3}x[A^{ik}\delta g_{ik} + B\delta \chi +
C_{ik}\delta \pi^{ik} + D\delta \pi_{\chi}]}
\end{equation}
whose coefficients $A^{ik}$, $B$, $C^{ik}$, D, when inserted into the
equations $\dot{\pi^{ik}} = - A^{ik}$, $\dot{g_{ik}} = C_{ik}$,
$\dot{\pi_{\chi}} = - B$, $\dot{\chi} = D$ yeld, together with the
constraints ${\cal{H}}\approx 0$ and ${\cal{H}}^{i} \approx 0$, the full
Einstein's equations for a general 4-geometry with a dust field. However,
when the spacelike hypersufaces are open, the variation of $H$ gives,
together with (\ref{deltaH}), the surface terms
\begin{equation}
\begin{array}{l}
- S_{T} = - \frac{1}{16\pi}\oint_{B}{d^{2}S_{l}G^{ijkl}(N\delta g_{ij \| k} -
N_{, k}\delta g_{ij})}\\
\\
- \oint_{B}{d^{2}S_{l}[2N_{k}\delta \pi^{kl} + (2N^{k}\pi^{jl} -
N^{l}\pi^{jk})\delta g_{jk}]}\\
\\
+  \oint_{B}{d^{2}S_{l}N\pi_{\chi}\chi^{, l}\delta \chi (m^{2} +
\chi_{, i}\chi^{, i})^{- 1/2}}
\\
\\
+ \oint_{B}{d^{2}S_{l}N^{l}\pi_{\chi}\delta \chi},
\label{st}
\end{array}
\end{equation}
where $d^{2}S_{l} = \frac{1}{2!}\epsilon_{ljk}dx^{j}\wedge dx^{k}$,
$\epsilon_{ljk}$ being the 3-dimensional metric independent totally
antisimetric object, B is a 2-sphere at infinity,
$G^{ijkl} = 1/2 g^{1/2}(g^{ik}g^{jl} + g^{il}g^{jk} -2g^{ij}g^{kl})$, and $g^{ik}$ is the inverse of $g_{ik}$ .

If the surface term (\ref{st}) does not vanish asymptotically, the functional
derivatives
of the hamiltonian cannot be defined due to the presence of them
in the expression of $\delta H$. The usual remedy to this problem is to evaluate
which surface terms in (\ref{st}) survive for specific asymptotic strutures,
and add to the hamiltonian (\ref{ham}) a suitable surface term such that, for
variations respecting
the asymptotic struture of the field, the variation of the total hamiltonian
$H_{T} = H +E$ yields
\begin{equation}
\begin{array}{l}
\delta H_{T} = \delta (H + E) = \delta H + \delta E =\\
=\int{d^{3}x[A^{ik}\delta g_{ik} + B\delta \chi + C_{ik}\delta \pi^{ik} +
D\delta \pi_{\chi}]} - S_{T} +  S_{T}
\end{array} 
\end{equation}
where E is constructed in such a way that $\delta E = S_{T}$. As a consequence,
the new hamiltonian $H_{T}$ has now well defined variations yielding the
correct Einstein's
equations. This procedure has been successfully applied to asymptotically
flat and anti-de Sitter spaces \cite{tei1,tei2}. We will now investigate if
this procedure can be done for asymptotically FRLW spacetime.

\section{Two Classes of Asymptotically FLRW Spacetime}

Our starting point is the class of Tolman metrics
\begin{equation}
\label{tolman}
ds^{2} = (1 - kr^{2})^{-1} R'^{2}dr^{2} + R^{2}(d\theta^{2} + {\sin}^{2}\theta
d\phi^{2}) - dt^{2},  
\end{equation}
where $r$, $\theta$ , $\phi$ are comoving coordinates with the spherically
symmetric fluid in the model, $R(r,t) \geq 0$, with $R(0,t) = 0$,
$r \geq 0$, $0 \leq \theta \leq \pi$ , $0 \leq \phi \leq 2\pi$,
a prime denotes $\frac{\partial}{\partial r}$, and $k = 0,-1$, as there are no
open models with k = 1.

If the line element (8) represents asymptotically FLRW spacetimes
then, asymptotically, one should have
\begin{equation}
R(r,t) = ra(t)\biggl[1 + \frac{f(t)}{r} + \frac{g(t)}{r^{2}} + ...\biggr]
\end{equation}
which implies that
\begin{equation}
R'(r,t) = a(t)\biggl[1 - \frac{g(t)}{r^{2}} + ...\biggr]
\end{equation}
Hence,
\begin{equation}
g_{22} =  {\stackrel{\circ}{g}}_{22}\biggl[1 + \frac{2f}{r} +
\frac{(f^{2} + 2g)}{r^{2}} + ...\biggr]
\end{equation}
\begin{equation}
g_{11} = {\stackrel{\circ}{g}}_{11}\biggl[1 - \frac{2g}{r^{2}} + ...\biggr],
\end{equation}
and $g_{33} = {\sin}^{2}\theta g_{22}$, where $ {\stackrel{\circ}{g}}_{ik}$
represents the background spacelike metric of the FLRW spacetime at infinity,
and we define $h_{ik} \equiv g_{ik} -  {\stackrel{\circ}{g}}_{ik}$.

For the momenta,
\begin{equation}
\pi^{ik} = - g^{1/2}(K^{ik} - g^{ik}K),
\end{equation}
where
\begin{equation}
K_{ik} = - \frac{1}{2N}(\dot{g_ {ik}} - N_{i \parallel k} - N_{k \parallel i}),
\end{equation}
and $K = g_{ik}K^{ik}$. According to the above expressions one has,
\begin{equation}
\label{mom1}
\pi^{ik} = {\stackrel{\circ}{\pi}}^{il}\biggl[\delta^{k}_{l} + \frac{b^{k}_{l}(t)}{r}
+ ...\biggr]
\end{equation}
where $ {\stackrel{\circ}{\pi}}^{ik}$ are the background momenta given by
\begin{equation}
 {\stackrel{\circ}{\pi}}^{ik} = - 2r^{2}{\sin}\theta \dot{a}a^{2}
 {\stackrel{\circ}{g}}^{ik}.
\end{equation}
There is a necessary contribuition to the $1/r$ dependence in (\ref{mom1}) coming
from the term $g^{1/2} \simeq  {\stackrel{\circ}{g}}^{1/2}[1 +
{\stackrel{\circ}{g}}^{ik}h_{ik} + ...]$. The precise form of $b^{k}_{l}(t)$ in terms
of $g(t)$ and $f(t)$ is not important for what follows.

For the dust field, as the velocity field of the backgound is
$ {\stackrel{\circ}{u}}^{\mu} = \delta^{\mu}_{0}$ then, asymptotically,
$\chi \simeq mt[1 + s(t)/r]$, $n \simeq
{\stackrel{\circ}{n}}(t)[1+ q(t)/r]$, and, from (\ref{velocity}) and (\ref{momd}),
one obtains
\begin{equation}
\pi_{\chi} = {\stackrel{\circ}{\pi}}_{\chi}(t)\biggl[1 + \frac{p(t)}{r}+...\biggr].
\end{equation}

To obtain the classes of asymptotically geometries which contain ({\ref{tolman})
and which are invariant under the action of the isometries of FLRW spacetimes
at infinity, one must take the asymptotic deviations of the metric
(\ref{tolman}), the field $\chi$ and
their momenta with respect to the reference FLRW space,
\begin{equation}
\begin{array}{rcl}
h_{ik} = g_{ik} -  {\stackrel{\circ}{g}}_{ik} &
q = \chi -  {\stackrel{\circ}{\chi}}
\\
p^{ik} = \pi^{ik} - {\stackrel{\circ}{\pi}}^{ik}  &
p_{\chi} = \pi_{\chi} - {\stackrel{\circ}{\pi}}_{\chi},
\label{var}
\end{array}
\end{equation}
and act on them with the Killing
vectors of FLRW spaces.
The Killing vectors of the homogeneous and isotropic hypersufaces of asymptotic
FLRW spacetimes read
\begin{equation}
\begin{array}{l}
\vec{\zeta}_{(12)} = \frac{\partial}{\partial{\phi}}\\
\\
\vec{\zeta}_{(23)} = - {\sin}\phi \frac{\partial}{\partial{\theta}} -
{{\rm cotg}} \theta {{\rm cos}} \phi \frac{\partial}{\partial \phi}\\
\\
\vec{\zeta}_{(31)} = {{\rm cos}} \phi \frac{\partial}{\partial{\theta}} -
{{\rm cotg}} \theta {\sin}\phi \frac{\partial}{\partial \phi}
\label{rot}
\end{array}
\end{equation}

\begin{equation}
\begin{array}{l}
\vec{\zeta}_{(14)} = (1 - kr^{2})^{1/2}{\sin}\theta {{\rm cos}}
\phi \frac{\partial}
{\partial{r}} + \frac{(1 - kr^{2})^{1/2}}{r}{{\rm cos}} \theta
{\cos}\phi \frac{\partial}
{\partial{\theta}} - \frac{(1 - kr^{2})^{1/2}}{r}\frac{{\sin}\phi}{{\sin}\theta}
\frac{\partial}{\partial{\phi}}\\
\\
\vec{\zeta}_{(24)} = (1 - kr^{2})^{1/2}{\sin}\theta {\sin}\phi
\frac{\partial}{\partial{r}} + \frac{(1 - kr^{2})^{1/2}}{r}{{\rm cos}} \theta
{\sin}\phi \frac{\partial}{\partial{\theta}} + \frac{(1 - kr^{2})^{1/2}}{r}
\frac{{{\rm cos}} \phi}{{\sin}\theta}\frac{\partial}{\partial \phi}\\
\\
\vec{\zeta}_{(34)} = (1 - kr^{2})^{1/2}{{\rm cos}} \theta \frac{\partial}
{\partial{r}} -
\frac{(1 - kr^{2})^{1/2}}{r}{\sin}\theta \frac{\partial}{\partial{\theta}}
\label{trans}
\end{array}
\end{equation}
The Killing vectors generators of rotations are written in (\ref{rot}), while
the generators of ``translations'' are written in (\ref{trans}).

The asymptotic strutures of
theses classes and the relevant surface terms in (\ref{st}) will depend on the
value of $k$, yielding two distinct classes of asymptotic behavior,
which will be treated separately.

\subsection{The k = 0 case}

 The background phase space variables are:
\begin{equation}
\begin{array}{rcl}
 {\stackrel{\circ}{g}}_{11} = a^{2}(t); &  {\stackrel{\circ}{g}}_{22} =
 a^{2}(t)r^{2};
 &  {\stackrel{\circ}{g}}_{33} = {\sin}^{2}\theta  {\stackrel{\circ}{g}}_{22}\\
\\
 {\stackrel{\circ}{\pi}}^{11} = -2r^{2}{\sin}\theta\dot{a}(t); &
 {\stackrel{\circ}{\pi}}^{22} = -2{\sin}\theta\dot{a}(t); &
 {\stackrel{\circ}{\pi}}^{33} =  {\stackrel{\circ}{\pi}}^{22}{\sin}^{-2}\theta\\
\\
 {\stackrel{\circ}{\chi}} = mt; &  {\stackrel{\circ}{\pi}}_{\chi}
 = na^3 r^{2}{\sin}\theta
\label{gfundo}
\end{array}
\end{equation}

The asymptotic deviations of (\ref{tolman}) from this background behaviour read
\begin{equation}
\begin{array}{rcl}
h_{11} \sim -2a^{2}(t)g(t)r^{-2}; & h_{22} \sim a^{2}(t)\{ 2r(t)f(t) +
[f^{2}(t) + 2g(t)]\}; \\
\\
p^{11} \sim n_{1}(t)r; & p^{22} \sim  n_{2}(t)r^{-1};\\
\\
h_{33} = {\sin}^{2}\theta h_{22}; & p^{33} =
p^{22}{\sin}^{-2}\theta;\\
\\
q \sim n_{3}(t)r^{-1}; & p_{\chi} \sim n_{4}(t)r,
\label{asymptvar}
\end{array}
\end{equation}
where the $n_i(t)$ are functions of time depending on the particular Tolman
solution we take.

Acting on these deviations with the Killing vectors of the k = 0 FLRW spacetimes
written in (\ref{rot}) and (\ref{trans}),
we obtain the general asymptotic behaviors:
\begin{equation}
\begin{array}{l}
h_{11} \sim r^{-2}l_{11}(t,\theta,\phi) + O(r^{-3})\\
\\
h_{22} \sim rl_{22}(t,\theta,\phi) + O(r^{0})\\
\\
h_{33} \sim rl_{33}(t,\theta,\phi) + O(r^{0})\\
\\
h_{12} \sim r^{-1}l_{12}(t,\theta,\phi) + O(r^{-2})\\
\\
h_{13} \sim r^{-1}l_{13}(t,\theta,\phi) + O(r^{-2})\\
\\
h_{23} \sim r^{-1}l_{23}(t,\theta,\phi) + O(r^{-2})\\
\\
p^{11} \sim rM^{11}(t,\theta,\phi) + O(r^{0})\\
\\
p^{22} \sim r^{-1}M^{22}(t,\theta,\phi) + O(r^{-2})\\
\\
p^{33} \sim r^{-1}M^{33}(t,\theta,\phi) + O(r^{-2})\\
\\
p^{12} \sim r^{-1}M^{12}(t,\theta,\phi) + O(r^{-2})\\
\\
 p^{13} \sim r^{-1}M^{13}(t,\theta,\phi) + O(r^{-2})\\
\\
p^{23} \sim r^{-1}M^{23}(t,\theta,\phi) + O(r^{-2})\\
\\
p_{\chi} \sim rM_{\chi}(t,\theta,\phi)  + O(r^{0})\\
\\
q \sim r^{-1}h_{\chi}(t,\theta,\phi) + O(r^{-2})
\label{asymptk0}
\end{array}
\end{equation}

The asymptotic behavior of the lapse function can be evaluated by demanding
that the dynamics do not spoil the above relations. The dynamics yields,
\begin{equation}
\label{dyn}
\dot{g}_{ik} = 2Ng^{-1/2}\biggl[\pi_{ik} - \frac{1}{2}g_{ik}\pi\biggr] +
N_{i \parallel k} + N_{k \parallel i}
\end{equation}
Evaluating it for i = k = 1, we find that $\delta N \sim O(r^{-2})$
for $\dot{g}_{11} \sim O(r^{-2})$ as $g_{11}$.

For the shift functions,
if one demands that the asymptotic hypersufaces be orthogonal to the fluid
4-velocity, then the hypersuface normal vector
$\eta^{\mu} = N^{-1}(1\,,N^{i}) \sim  u^{\mu} =
- g^{\mu\nu}\chi_{, \nu}/m$ which implies from (\ref{var}), (\ref{gfundo}) and
(\ref{asymptvar}) that $N^{1} \sim O(r^{-2})$, and $N^{2} \sim N^3 \sim O(r^{-3})$.

\subsection{The k = -1 case}
The background phase space variables are now given by:
 \begin{equation}
\begin{array}{rcl}
 {\stackrel{\circ}{g}}_{11} = a^{2}(t)(1 + r^{2})^{-1}; &
 {\stackrel{\circ}{g}}_{22} = a^{2}(t)r^{2}; \\
\\
 {\stackrel{\circ}{\pi}}^{11} = -2r^{2}(1 + r^{2})^{1/2}{\sin}\theta\dot{a}(t); &
 {\stackrel{\circ}{\pi}}^{22} = -2{\sin}\theta\dot{a}(t)(1 + r^{2})^{-1/2}; \\
 \\
 {\stackrel{\circ}{g}}_{33} =
 {\sin}^{2}\theta  {\stackrel{\circ}{g}}_{22}; & 
 {\stackrel{\circ}{\pi}}^{33} =  {\stackrel{\circ}{\pi}}^{22}{\sin}^{-2}\theta\\
\\
 {\stackrel{\circ}{\chi}} = mt; &  {\stackrel{\circ}{\pi}}_{\chi}
 = n a^3 r^{2}{\sin}\theta (1 + r^{2})^{-1/2}
\end{array}
\end{equation}

The asymptotic deviations of (\ref{tolman}) from this background now read
\begin{equation}
\begin{array}{rcl}
h_{11} \sim -2a^{2}(t)g(t)r^{-4};
& h_{22} \sim a^{2}(t)\{2rf(t) + [f^{2}(t) + 2g(t)]\}; \\
\\
p^{11} \sim n_{1}(t)r; & p^{22} \sim  n_{2}(t)r^{-1}; \\
\\
h_{33} = {\sin}^{2}\theta h_{22}; & p^{33} = p^{22}{\sin}^{-2}\theta;\\
\\
q \sim n_{3}(t)r^{-1}; & p_{\chi} \sim n_{4}(t).
\label{g-1}
\end{array}
\end{equation}

The action of the Killing vectors of the k = -1 FLRW spacetimes listed
in (\ref{rot}) and (\ref{trans}) yields the following general behavior,
\begin{equation}
\begin{array}{l}
h_{11} \sim r^{-4}l_{11}(t,\theta,\phi) + O(r^{-5})\\
\\
h_{22} \sim rl_{22}(t,\theta,\phi) + O(r^{0})\\
\\
h_{33} \sim rl_{33}(t,\theta,\phi) + O(r^{0})\\
\\
h_{12} \sim r^{-2}l_{12}(t,\theta,\phi) + O(r^{-3})\\
\\
h_{13} \sim r^{-2}l_{13}(t,\theta,\phi) + O(r^{-3})\\
\\
h_{23} \sim r^{-1}l_{23}(t,\theta,\phi) + O(r^{-2})\\
\\
p^{11} \sim r^{2}M^{11}(t,\theta,\phi) + O(r)\\
\\
p^{22} \sim r^{-2}M^{22}(t,\theta,\phi) + O(r^{-3})\\
\\
p^{33} \sim r^{-2}M^{33}(t,\theta,\phi) + O(r^{-3})\\
\\
p^{12} \sim r^{-1}M^{12}(t,\theta,\phi) + O(r^{-2})\\
\\
 p^{13} \sim r^{-1}M^{13}(t,\theta,\phi) + O(r^{-2})\\
\\
p^{23} \sim r^{-1}M^{23}(t,\theta,\phi) + O(r^{-2})\\
\\
p_{\chi} \sim M_{\chi}(t,\theta,\phi)  + O(r^{-1})\\
\\
q \sim r^{-2}h_{\chi}(t,\theta,\phi) + O(r^{-3})
\label{k-1}
\end{array}
\end{equation}
where the last two behaviors come from the fact that $u^{i}$,
according to (3), must be zero at infinity.

Using the same arguments as in the
k = 0 case, we obtain that $N \approx 1 + O(r^{-2})$,
$N^{1} \approx O(r^{-1})$,
 $N^{2} \approx O(r^{-4})$, $N^{3} \approx O(r^{-4})$.

\section{The Hamiltonians of Two Classes of Asymptotically
FLRW Spacetimes}

We are now in position of evaluating the surface integrals (\ref{st}).
In spherical coordinates, $d^{2}S_{l} = d\theta d\phi \delta^{1}_{l}$.

\subsection{The k = 0 case}
The unique terms on the integrand which may not vanish asymptotically are:
\begin{equation}
- S_{T} = -  {\stackrel{\circ}{g}}^{1/2} {\stackrel{\circ}{g}}^{11}
[ {\stackrel{\circ}{g}}^{22}
{\stackrel{\circ}{\Gamma}}^{2}_{12}\delta g_{22} +
{\stackrel{\circ}{g}}^{33} {\stackrel{\circ}{\Gamma}}^{3}_{13}\delta g_{33} +
{\stackrel{\circ}{g}}^{22}\delta g_{22 ; 1} +
{\stackrel{\circ}{g}}^{33}\delta g_{33 ; 1}],
\end{equation}
where the semicolon means the covariant derivative with respect to the
background metric. However, as $\delta g_{22} \approx \delta g_{33}
\approx {\rm O}(r)$,
and as $ {\stackrel{\circ}{\Gamma}}^{2}_{12} =
{\stackrel{\circ}{\Gamma}}^{3}_{13} = r^{-1}$,
the terms in the above expression cancel out at this order.
Hence $S_{T} = 0$ for k = 0 and it is not necessary to suplement the hamiltonian
with any surface term. It means that, on shell, $H = 0$.

\subsection{The k = -1 case}

The surface term (\ref{st}) can now be calculed. The relevant terms are:
\begin{equation}
\begin{array}{l}
-S_{T} = -\frac{1}{16\pi}\int_{B}{d^{2}S_{1} {\stackrel{\circ}{g}}^{1/2}
{\stackrel{\circ}{g}}^{11} {\stackrel{\circ}{g}}^{22}[\delta g_{21 , 2}
-  {\stackrel{\circ}{\Gamma}}^{1}_{22}\delta g_{11}}\\
\\
- ( {\stackrel{\circ}{\Gamma}}^{2}_{12} +  \Delta\Gamma^{2}_{12})
 (\delta g_{22} + \delta  {\stackrel{\circ}{g}}^{22}) - \delta g_{22 , 1}]
 \\
\\
-\frac{1}{16\pi}\int_{B}{d^{2}S_{1} {\stackrel{\circ}{g}}^{1/2}
{\stackrel{\circ}{g}}^{33}
{\stackrel{\circ}{g}}^{11}[\delta g_{31 , 3} -
{\stackrel{\circ}{\Gamma}}^{1}_{33}\delta g^{11}}\\
\\
- ({\stackrel{\circ}{\Gamma}}^{3}_{13} +  \Delta\Gamma^{3}_{13})(\delta g_{33} +
\delta  {\stackrel{\circ}{g}}^{33}) - \delta g_{33 , 1}] ,
\end{array}
\end{equation}
where $\Delta\Gamma^{2}_{12}=1/2({\stackrel{\circ}{g}}^{22}h_{22,1}+
h^{22}{\stackrel{\circ}{g}}_{22,1})$ and
$\Delta\Gamma^{3}_{13}=1/2({\stackrel{\circ}{g}}^{33}h_{33,1}+
h^{33}{\stackrel{\circ}{g}}_{33,1})$ are the
next leading order terms of $\Gamma^{2}_{12}$ and
$\Gamma^{3}_{13}$ after ${\stackrel{\circ}{\Gamma}}^{2}_{12}$ and
${\stackrel{\circ}{\Gamma}}^{3}_{13}$, respectively.
The divergent terms cancel out ($ {\stackrel{\circ}{\Gamma}}^{2}_{12} =
{\stackrel{\circ}{\Gamma}}^{3}_{13} = r^{-1}$) and this surface term is finite.
The terms containing the deviations $\Delta\Gamma$ of the connections
with respect to the background connection, which are of order $r^{-2}$,
can be written as
\begin{equation}
\frac{1}{16\pi}\delta \int{d^{2}S_{l}\frac{1}{8}(h^{ij}h_{ij})_{; m}
{\stackrel{\circ}{g}}^{ml} {\stackrel{\circ}{g}}^{1/2}}.
\end{equation}
This can be proved by using that, at leading order,
$h_{22}$, $\delta g_{22}$, $h_{33}$, $\delta g_{33}$ are of order r and hence
$h_{22}\delta g_{22,1} = h_{22,1}\delta g_{22}$ and
$h_{33}\delta g_{33,1} = h_{33,1}\delta g_{33}$.
Then the total no null surface terms for this k = -1 case reads
\begin{equation}
 - S_{T} = - \frac{1}{16\pi}\delta \int{d^{2}S_{l}
 \biggl[ {\stackrel{\circ}{G}}^{ijkl}h_{ij ; k} -
 \frac{1}{8}(h^{ij}h_{ij})_{; m} {\stackrel{\circ}{g}}^{ml}
 {\stackrel{\circ}{g}}^{1/2}\biggr]}
\end{equation}
The total hamiltonian with well defined functional derivatives is
$H_{T} = H + E$, where
\begin{equation}
 E = \frac{1}{16\pi}\int{d^{2}S_{l}[ {\stackrel{\circ}{G}}^{ijkl}h_{ij ; k} -
 \frac{1}{8}(h^{ij}h_{ij})_{; m} {\stackrel{\circ}{g}}^{ml}
 {\stackrel{\circ}{g}}^{1/2}]}.
\label{E-1}
\end{equation}

We would like to emphasize once again that the variations in (\ref{E-1})
are for fixed
background at infinity, which means that $\delta  {\stackrel{\circ}{g}}_{ml} = 0$
at infinity. The surface term $E$ is completely covariant under general
spatial coordinate transformations and finite.

For the case of Tolman metrics, where the asymptotic deviations from FLRW are
given in (\ref{g-1}), the quantity $E$ given in (\ref{E-1}) is identically zero
because of cancellation of its terms.
This happens due to the three conditions satisfied by the Tolman metric
(\ref{tolman}):

i) The functions $f$ and $g$ in $h_{11}$ and $h_{22}$ depend only on time.

ii) The function $g(t)$ appearing in $h_{11}$ is the same as the one appearing
in the second term of $h_{22}$.

iii) $h_{33}= {\sin}^{2}\theta h_{22}$.

If one of these conditions is not satisfied, the term $E$ may not be null.

\section{Conclusion}

We have constructed consistent hamiltonians for some classes of
asymptotically FLRW spacetimes.

In the case $k=0$, there is no need
to add surface terms to the usual hamiltonian of GR in order for
the hamiltonian formalism be well defined.

In the case $k=-1$, a surface term  must be present which is shown in
Eq. (\ref{E-1}). For Tolman geometries given in (\ref{tolman}),
this surface term is zero, although it should not be zero in more general
cases.

If we interpret the value of the hamiltonian evaluated in a solution of
Einstein's equations as the ``energy'' of this solution with respect to
the FLRW spacetime they approach at infinity (its reference spacetime), which
is not
necessarily conserved because of the absence of timelike Killing vectors in
such solutions, then we can say that any asymptotic FLRW spacetime
with $k=0$ and Tolman solutions with $k=-1$ have the same ``energy'' as
their reference FLRW spacetimes.

Note that the terms involving the dust field degrees of freedom do
not contribute to the surface terms in any case. The same must be true
for other perfect fluids. A cosmological constant also does not contribute
because it appears in the hamiltonian without derivative terms ($g^{1/2}\Lambda$).

One important consequence of these results is that if one wishes to study
the midisuperspace quantization of geometries with asymptotic behavior given
in (\ref{k-1}), then the surface term (\ref{E-1}) must be taken into
account, yielding a non trivial Schr\"{o}dinger equation besides the Wheeler-DeWitt
equation. For the $k=0$ case, the situation is the same as in midisuperspace
quantization of closed geometries: the only relevant equation is the
Wheeler-DeWitt equation as long as, in this case, there are also no surface terms.

It should be interesting to evaluate
the surface term (\ref{E-1}) for asymptotic FLRW spacetimes satisfying
(\ref{k-1}) which are different from Tolman geometries, and study the physical
significance of this term in this geometry. This will be the subject of our
further investigations.

\section*{ACKNOWLEDGMENTS}

We would like to thank the Cosmology Group of
CBPF for useful discussions, and CNPq of Brazil for financial support.

\end{document}